\documentclass[aps,prb,twocolumn,showpacs,superscriptaddress]{revtex4}
\usepackage{graphicx}

\begin{document}

\title{Evolution of the low energy spin dynamics in 
electron-doped high-transition temperature 
superconductor  Pr$_{0.88}$LaCe$_{0.12}$CuO$_{4-\delta}$}

\author{Stephen D. Wilson}
\affiliation{
Department of Physics and Astronomy, The University of Tennessee, Knoxville, Tennessee 37996-1200, USA
}
\author{Shiliang Li}
\affiliation{
Department of Physics and Astronomy, The University of Tennessee, Knoxville, Tennessee 37996-1200, USA
}
\author{Pengcheng Dai}
\email{daip@ornl.gov}
\affiliation{
Department of Physics and Astronomy, The University of Tennessee, Knoxville, Tennessee 37996-1200, USA
}
\affiliation{
Center for Neutron Scattering, Oak Ridge National Laboratory, Oak Ridge, Tennessee 37831-6393, USA}
\author{Wei Bao}
\affiliation{
Los Alamos National Laboratory, Los Alamos, New Mexico 87545, USA}
\author{Jae-Ho Chung}
\affiliation{
NIST Center for Neutron Reseatch, National Institute of Standards and Technology, Gaithersburg, Maryland 20899-8562, USA
}
\affiliation{
Department of Materials Science and Engineering, University of Maryland, College Park, Maryland 20742-2115, USA
}
\author{H. J. Kang}
\affiliation{
NIST Center for Neutron Reseatch, National Institute of Standards and Technology, Gaithersburg, Maryland 20899-8562, USA
}
\affiliation{
Department of Materials Science and Engineering, University of Maryland, College Park, Maryland 20742-2115, USA
}
\author{Seung-Hun Lee}
\affiliation{
Department of Physics, University of Virginia, Charlottesville, Virginia 22904-4714 USA
}
\author{Seiki Komiya}
\affiliation{
Central Research Institute of Electric Power Industry, Komae, Tokyo 201-8511, Japan
}
\author{Yoichi Ando}
\affiliation{
Central Research Institute of Electric Power Industry, Komae, Tokyo 201-8511, Japan
}
\author{Qimiao Si}
\affiliation{
Department of Physics and Astronomy, Rice University, Houston, Texas 77005 USA
}

\begin{abstract}
We use inelastic neutron scattering to explore the evolution of the low energy spin dynamics in the electron-doped cuprate Pr$_{0.88}$LaCe$_{0.12}$CuO$_{4-\delta}$ (PLCCO) as the system is tuned from its nonsuperconducting, as-grown 
antiferromagnetic (AF) state into an optimally-doped superconductor ($T_c\approx24$ K) without static AF order.  The low temperature, low energy response of the spin excitations in under-doped samples is coupled to the presence of the AF phase, whereas the low-energy magnetic response for samples near optimal $T_c$ exhibits spin fluctuations surprisingly insensitive to the sample temperature.  This evolution of the low energy excitations is consistent with the influence of a quantum critical point in the phase diagram of PLCCO associated with the suppression of
the static AF order. We carried out scaling analysis of the data and discuss
the influence of quantum critical dynamics in the observed excitation spectrum.       
\end{abstract}


\maketitle

\section{Introduction}

The families of high-transition-temperature (high-$T_c$) superconductors are fundamentally composed of two-dimensional copper oxygen planes into which charge carriers, either holes or electrons, are doped.  Prior to doping charge carriers, the parent compounds of the high-$T_c$ copper oxides are antiferromagnetically ordered insulators whose spin dynamics are well modeled by a two-dimensional Heisenberg antiferromagnetic (AF) formalism with an anomalously large nearest-neighbor exchange coupling ($J > 100$ meV) between the Cu sites within the CuO$_2$ plane \cite{coldea, hayden96, bourges97}.  As spin fluctuations may play a crucial role in the mechanism of high-$T_c$ superconductivity \cite{scalapino,dai99,woo}, it is imperative to have a comprehensive picture on 
how the spin dynamics of the undoped AF parent compounds evolve as they are tuned towards optimally-doped superconductivity.  While a comprehensive picture of this carrier induced modification to the spin dynamics has  emerged for different classes of 
hole-doped high-$T_c$ materials  \cite{woo,hayden04,tranquada04,christen04,tranquada05}, experiments exploring spin fluctuations in electron-doped copper oxides are just beginning \cite{yamada}. As a consequence, studies of  electron-doped materials provide a unique litmus for testing the electron-hole symmetry in spin dynamical
properties.  If spin fluctuations are fundamental to the mechanism of high-$T_c$ superconductivity, they should
have universal features for all copper-oxide systems.

In the case of hole-doped high-$T_c$ superconductors such as 
YBa$_2$Cu$_3$O$_{6+x}$ (YBCO), the most prominent feature in its spin excitation spectrum is a sharp 
magnetic excitation termed `resonance' observed by inelastic neutron scattering.  The resonance is 
centered at the AF ordering wavevector ${\bf Q}=(1/2,1/2)$ in the two-dimensional reciprocal space 
of the CuO$_2$ planes (see inset
of Fig. 1a) and is intimately related to superconductivity \cite{dai99,woo,hayden04}.   
Our recent discovery of the resonance mode in the electron-doped 
Pr$_{0.88}$LaCe$_{0.12}$CuO$_{4-\delta}$ ($T_c=24$ K) demonstrates that the
resonance is a universal feature of high-$T_c$ 
copper oxides 
regardless of carrier type \cite{wilson}.  This unifying feature in the spin excitation spectrum of both hole- and electron-doped cuprates contrasts the 
prevalent asymmetry known between the commensurate low energy spin fluctuations centered at 
${\bf Q}=(1/2,1/2)$ in the electron-doped cuprate systems \cite{yamada,wilson,wilson06} and the observed incommensurate spin excitations at ${\bf Q}=(0.5\pm\delta,0.5\pm\delta)$ in several classes of 
hole-doped materials \cite{tranquada04,yamada2,dai01}.  While the doping dependence of the resonance excitation has been observed to follow $E_r\approx5.8k_BT_c$ regardless of carrier type \cite{wilson} and the doping evolution of the incommensurate spin fluctuations in hole-doped materials is also controlled by $T_c$ \cite{yamada2,dai01}, the doping dependence of the low energy commensurate spin fluctuations 
in electron-doped materials remains unexplored.  

\begin{figure}
\includegraphics[scale=.40]{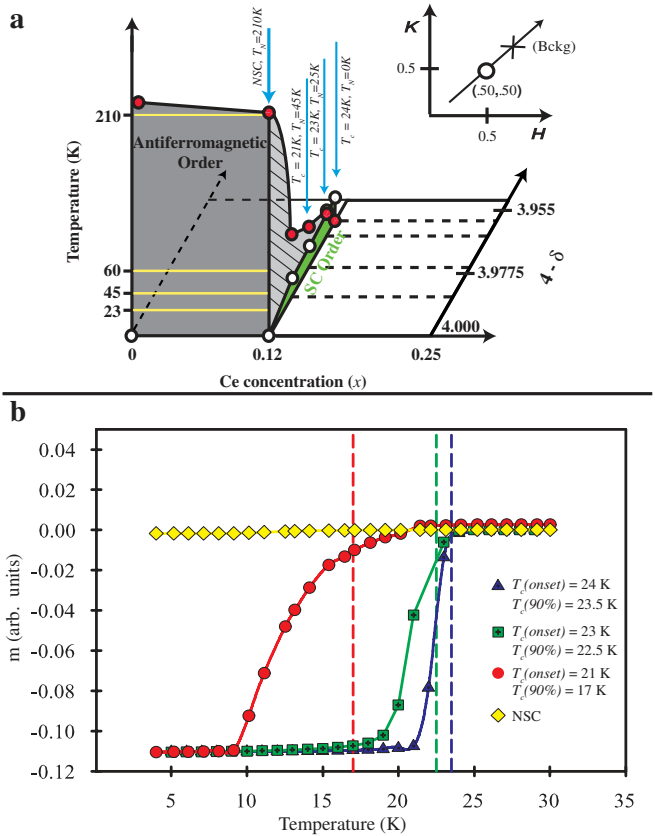}
\caption{
 Phase diagram of Pr$_{0.88}$LaCe$_{0.12}$CuO$_{4-\delta}$ (PLCCO) and magnetic bulk susceptibility measurements for PLCCO samples studied. a) Phase diagram of PLCCO from Kang {\it et al.} \cite{kang05} with the $T_c=23$ K sample added.  The blue arrows denote the sample compositions and locations on the phase diagram that were probed in the neutron experiments.  The inset shows the direction of $Q$-scans with the circle denoting $(0.5,0.5,0)$ and the cross indicating where the offset background was measured \cite{note1}. b) Bulk susceptibility measurements showing the superconducting phase transitions for the samples studied.  Both the onset temperature of superconductivity and the point of 90$\%$ of the normal state response (shown as dashed lines) are reported.  Susceptibilities have been normalized to one another for plotting purposes.  
 }
\end{figure}

For this reason, we chose to investigate the evolution of low frequency spin fluctuations in the electron-doped cuprate Pr$_{0.88}$LaCe$_{0.12}$CuO$_{4-\delta}$ (PLCCO).  
Compared with the prototypical electron-doped copper oxide Nd$_{2-x}$Ce$_x$CuO$_{4-\delta}$ (NCCO), 
studying spin fluctuations in PLCCO offers several distinct advantages: First,  
crystalline electric field (CEF) levels of 
Pr$^{3+}$ in the tetragonal unit cell of PLCCO have a 
nonmagnetic, singlet ground state, which avoids complications arising from
the magnetic ground state of Nd$^{3+}$ in NCCO \cite{boothroyd,sumarlin}. Second, PLCCO can be tuned from
an as-grown nonsuperconducting (NSC) antiferromagnet 
to an optimally doped superconductor, without static AF order, through an 
annealing process \cite{fujita,dai05,kang05}; whereas static AF order coexists with superconductivity in
NCCO even at optimal doping \cite{yamada,uefuji,kang03}. Finally, the cubic (Pr,La,Ce)$_2$O$_3$ impurity
phase arising from the annealing process \cite{kang05} in PLCCO has a nonmagnetic ground state as compared
to the magnetic ground state of the (Nd,Ce)$_2$O$_3$ impurity phase in NCCO \cite{matsuura03}. 
In this article, we present a comprehensive study of the low energy spin dynamics in PLCCO from 0.5 meV $\leq\hbar\omega \leq$ 5 meV in a variety of samples as PLCCO is tuned from a NSC antiferromagnet into an optimally
doped superconductor ($T_c=24$ K) through the annealing process \cite{dai05,kang05}.

Electron-doped copper oxides differ from the their hole-doped counterparts in that they require a post-growth annealing treatment in a low-oxygen atmosphere to remove excess oxygen and 
achieve superconductivity \cite{tokura}.  This means that 
the phase diagram of electron-doped copper oxides is three-dimensional as a function of both Ce and oxygen 
concentration (Fig. 1a). There are therefore   
two distinct ways to traverse the phase diagram of PLCCO: 
samples can be prepared at a fixed annealing condition but grown with variable Ce doping levels \cite{fujita}, or alternatively, samples can be grown with a fixed Ce concentration but with variable 
annealing treatment \cite{dai05,kang05}.  For our studies, we utilize the latter method with a fixed Ce concentration of $x=0.12$ and the resulting phase diagram is shown in Fig. 1a.  It is clear that the annealing
process induces a rapid suppression of the long-range AF order and the eventual emergence of a superconducting phase transition.  With continued oxygen removal, the superconducting phase is enhanced to an optimum $T_c$ where the static AF order is completely suppressed.  For samples which exhibit a superconducting phase, an additional quasi-two-dimensional spin density wave (SDW) order appears at the disallowed three-dimensional AF ordering wave vector 
$\bf{Q}$$=(1/2, 1/2)$ in the CuO$_2$ plane \cite{dai05,kang05}.  The SDW order has an onset of approximately $T_{N}$ for underdoped superconducting PLCCO samples.       
	
The rest of this article is organized as follows: The next section describes the experimental
procedure, including details of sample preparation and neutron spectrometer setup. In section III, we discuss 
low energy magnetic excitations in the as-grown, AF ordered NSC PLCCO.  
Subsequent sections (IV-VI) cover spin fluctuations in superconducting samples annealed gradually toward optimal superconductivity.  As AF order is suppressed in the system, the low energy excitations transition from regimes coupled to the onset of the AF phase into a virtually temperature independent regime similar to those observed in several heavy Fermion systems known to be near a quantum critical point (QCP) in the phase diagram 
\cite{schroeder,aronson,wilson05,stewart}. In section VII, we discuss the applicability of quantum critical scaling in describing the spin dynamics of PLCCO with different transition temperatures.
Finally, in section VIII we briefly summarize the key conclusions of the work.

\section{Experimental}
For our experiments, we grew high quality single crystal PLCCO samples in an infrared mirror image furnace using the traveling solvent floating zone technique.  All samples were confirmed to have a mosaic of $<1^\circ$.  Following their growth, samples were annealed in either a high vacuum environment ($P<10^{-6}$ mbar) or in an argon gas environment at variable temperatures.  The resulting magnetic susceptibility and superconducting phase transition were measured in a SQUID magnetometer and are shown for each sample in Fig. 1b.  The superconducting transition temperature is indicated by both the onset of the diamagnetic signal, $T_c$ (onset), and through the dashed lines showing the 90\% position of normal-state bulk susceptibility, $T_c$ (90\%).  For the remainder of the paper, we will reference these samples by their respective onset temperature of superconductivity, $T_c$ (onset).  The as-grown sample, with a mass of $\sim$3.0 g, is nonsuperconducting and is labeled as NSC.  The masses of the other PLCCO's are $\sim$1.8 g, $\sim$4.8 g, and $\sim$3.0 g for the $T_c=21$ K, $T_c=23$ K, and $T_c=24$ K samples, respectively.  An argon annealing atmosphere was used in treating the $T_c=21$ K sample at 940 $^\circ$C 
for 24 hours \cite{dai05}.  The other two samples were annealed for four days in a high vacuum environment at 765 $^\circ$C and 775 $^\circ$C for the $T_c=23$ K, and $T_c=24$ K samples, respectively.  

\begin{figure}
\includegraphics[scale=.40]{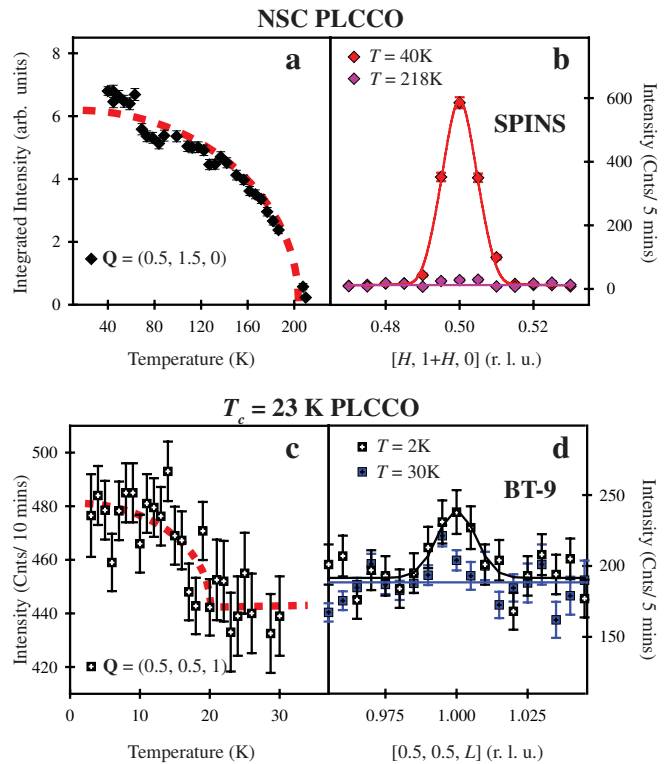}
\caption{
 Magnetic order parameters and AF Bragg peaks for both the NSC and $T_c=23$ K samples.  a) Order parameter for the N${\rm \acute{e}}$el phase in the NSC PLCCO.  All points are integrated intensities of the {\bf{Q}}=(0.5, 1.5, 0) AF Bragg reflection. b) Representative elastic $Q$-scans through {\bf{Q}}=(0.5, 1.5, 0) along the 
 $[H, 1+H, 0]$ direction both below $T_N$ at 40 K and above $T_N$ at 218 K. c) Magnetic AF order parameter for the $T_c=23$ K sample. The sample is aligned in the $[H,H,L]$ scattering zone.  
 Points are the peak intensities collected at the ${\bf{Q}}=(0.5, 0.5, 1)$ Bragg position.  d) Elastic $Q$-scans through the $(0.5, 0.5, 1)$ position along the $[0.5, 0.5, L]$ direction.  
 A weak AF Bragg reflection appears below $T_N$ at 2 K and disappears above $T_N$ at 30 K.    
 }
\end{figure}

We reference positions in reciprocal space at wave vector {\bf{Q}}$=(q_x, q_y, q_z)$ 
in \AA$^{-1}$ 
using $(H, K, L)$ (r.l.u.) notation, where $(H, K, L)=({q_x a}/2\pi,{q_y b}/2\pi,{q_z c}/2\pi)$ for the
tetragonal PLCCO unit cell (space group: $I4/mmm$ $a=b=3.98$ \AA, $c=12.27$ \AA ).  Neutron scattering experiments were performed at the NIST Center for Neutron Research on the SPINS and BT-9 triple-axis spectrometers.  Data was collected on the cold-neutron spectrometer SPINS with a fixed final energy of $E_f=5.0$ meV and collimations of open-80$^\prime$-sample-80$^\prime$-open-detector for NSC, $T_c=21$ K, and $T_c=23$ K samples, while an $E_f=3.7$ meV with identical collimations was used for SPINS experiments on the $T_c=24$ K samples.  For data collected from BT-9, an $E_f=14.7$ meV was used with collimations of 40$^\prime$-60$^\prime$-sample-80$^\prime$ -open-detector. Unless otherwise stated, PLCCO samples were aligned within the $[H,K,0]$ scattering plane for experiments on SPINS, and within the $[H,H,L]$ scattering zone for BT-9.  Alignment in the $[H,K,0]$ scattering zone allows for an effective integration along the $c$-axis of the quasi-two-dimensional, inelastic scattering from the Cu spins in uncorrelated CuO$_2$ planes (through relaxed out-of-plane resolution), whereas alignment in the $[H,H,L]$ scattering zone allows an accurate determination of the onset of the three-dimensional AF phase in the system.  Masses for the samples used in these neutron experiments correspond to the masses stated above in the bulk magnetization measurements with the exception of experiments probing the $T_c=24$ K system.  For experiments on this optimally-doped PLCCO system, a set of three $T_c=24$ K samples were co-aligned with a 
total mass of $\sim$9 g \cite{wilson}.  All samples were mounted and loaded into a liquid-He cooled cryostat, and the experiments were performed in the range from $T=2$ K  to $220$ K.

\begin{figure}
\includegraphics[scale=.42]{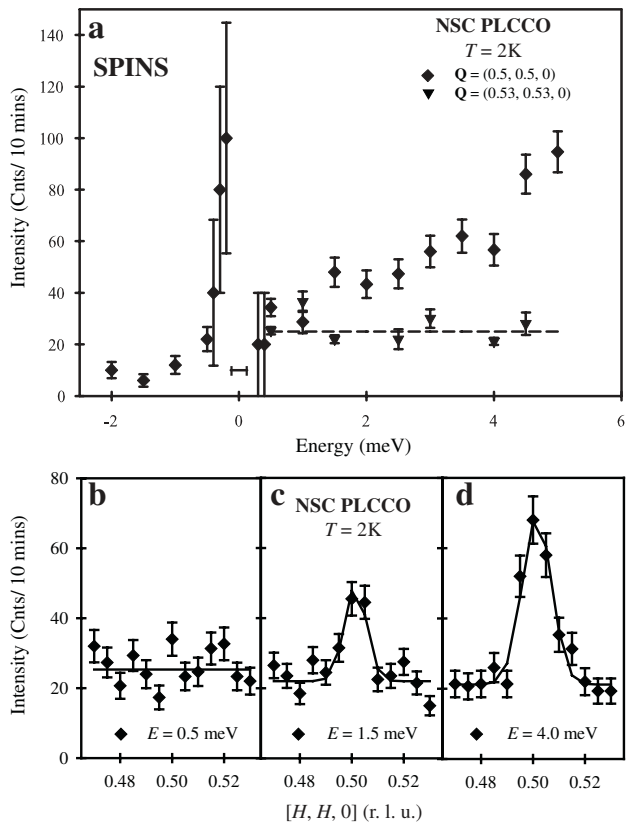}
\caption{
 Low energy spin excitations in NSC PLCCO. a) Constant-$Q$ scans taken at $T=2$ K. The peak signal was collected at ${\bf{Q}}=(0.5, 0.5, 0)$ (black diamonds) and the nonmagnetic background was measured at ${\bf{Q}}=(0.53, 0.53, 0)$ (black triangles).  The dashed line is a linear fit to the energy dependence of the nonmagnetic background.  The instrumental energy resolution of FHWM$=0.259$ meV is determined from vanadium scans and is shown as a solid bar under the elastic incoherent peak. b-d) Constant-$E$ scans along the $[H, H, 0]$ direction 
at $\hbar \omega =0.5$, 1.5, and 4.0 meV and collected at $T=$2 K.  Peaks are fitted by Gaussians centered at ${\bf Q}=(0.5, 0.5, 0)$ and are resolution-limited.            
 }
\end{figure}

\section{As-grown, NSC PLCCO}

We first opted to study the behavior of as-grown PLCCO, prior to any annealing treatment.  For this untreated system, long-range AF order persists up to a $T_N=210$ K as shown in Fig. 2a.  The rapid intensity
increase below about 80 K arises from the induced Pr$^{3+}$ moment through Cu$^{2+}$-Pr$^{3+}$ 
interaction \cite{lavrov04}. Scans through the allowed three-dimensional AF ordering wave vector $\bf{Q}$$=(0.5, 1.5, 0)$ show the appearance of a resolution-limited AF Bragg reflection centered at $\bf{Q}$$=(0.5, 1.5, 0)$ which diminishes above $T_N$ (Fig. 2b).  Due to the noncollinear structure of the zero field spin arrangement in PLCCO, the in-plane AF Bragg reflection at ${\bf Q}=(0.5, 0.5, 0)$ is disallowed, resulting in a gapped excitation spectrum at this wave vector \cite{kang05,petrigrand}.  Constant-$E$ scans along the 
$[H,H,0]$ direction show this low energy gap in Figs. 3b-d.  In Fig. 3b, ${\bf Q}$ scans at $T=2$ K through the ${\bf Q}=(0.5, 0.5, 0)$ position at $\hbar \omega =0.5$ meV show no magnetic scattering, whereas for $\hbar \omega \geq1.5$ meV clear resolution-limited spin-wave peaks are observed centered at ${\bf Q}=(0.5, 0.5, 0)$.  These resolution-limited spin-wave peaks are well fit by Gaussians and give minimum in-plane correlation lengths of $\xi _{min}=248\pm 40$ \AA\ at $\hbar \omega =1.5$ meV and $\xi _{min}=187\pm 21$ \AA\  
at $\hbar \omega =4.0$ meV \cite{kang05}.  Now plotting in Fig. 3a the constant-$Q$ scans at the 
${\bf Q}=(0.5, 0.5, 0)$ position, this low energy gap becomes clearer.  Background points taken at ${\bf Q}=(0.53, 0.53, 0)$ are overplotted with raw data taken at the peak ${\bf Q}=(0.5, 0.5, 0)$ position, showing the presence of the low energy gap to be $E_{gap}\approx 1.25$ meV.  This gap energy along with the enhancement in the observed magnetic scattering for $\hbar \omega >5$ meV is consistent with previous studies of the parent compound Pr$_2$CuO$_2$ in which the opening of an in-plane anisotropy gap was observed in addition to the lower interplane gap \cite{sumarlin,petrigrand}.

The temperature dependence of the low energy spin waves in this NSC sample is shown in Figs. 4a-c.  For energies above the gap, the spin-wave excitations increase in intensity with increasing temperature (for $T<T_N$) following the Bose population factor $[n(\omega )+1] = 1/(1-exp(-\hbar \omega/{k_B T}))$.  Removing the nonmagnetic background contributions to the scattering, we obtain $S($$\bf{Q}$$, \omega)$ and determine the imaginary part of the dynamic susceptibility, $\chi^{\prime \prime}($$\bf{Q}$$, \omega)$, using $S($$\bf{Q}$$, \omega) \propto [n(\omega )+1]\chi^{\prime \prime}({\bf Q},\omega)$.  Now plotting $\chi^{\prime \prime}({\bf Q},\omega)$ in Figs. 4d-f, the $T$-independence of the dynamic susceptibility for $E> E_{gap}$ becomes clear.  The peaks in $\chi^{\prime \prime}({\bf Q}, \omega)$ at these energies remain $T$-independent consistent with the excitations simply following the Bose statistics expected for spin-wave excitations.  The presence of bose-populated spin wave excitations in this NSC sample is similar to those observed in parent compounds La$_2$CuO$_4$ and Pr$_2$CuO$_4$ \cite{hayden96, bourges97}.  Above $T_{N}$, however, the low energy spin gap closes and spin fluctuations appear in the $\hbar\omega=0.5$ meV channel.  This is most likely due to the emergence of classical critical fluctuations arising from the suppression of AF order at $T= 210$ K, or due to a crossover from three-dimensional to two-dimensional spin fluctuations as the weak out-of-plane Cu exchange coupling breaks down above $T_N$.  The overall picture of spin dynamics in this NSC sample is well described by spin waves arising from the long-range AF order in this system, which now provides a baseline for studying how these spin excitations evolve as the system is tuned into superconductivity.

\begin{figure}
\includegraphics[scale=.40]{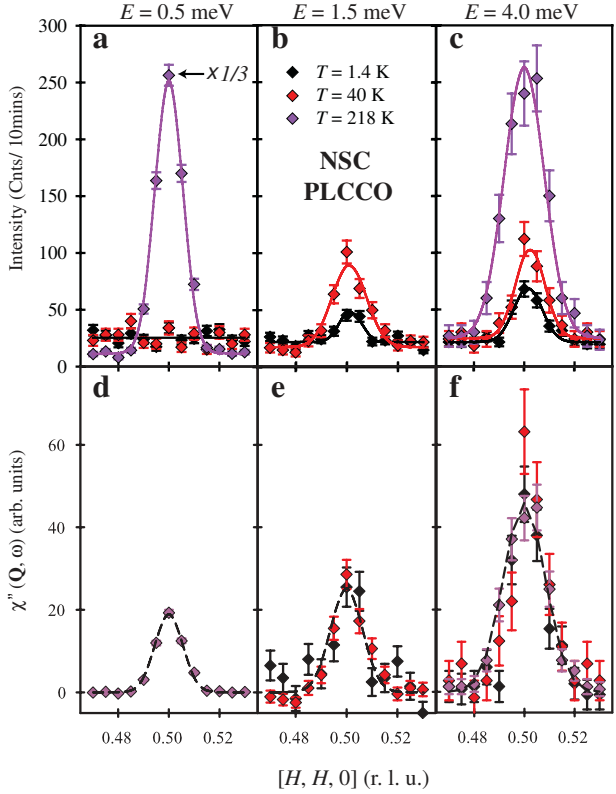}
\caption{
Temperature dependence of the inelastic neutron scattering intensity and the dynamic susceptibility in NSC PLCCO.  a-c) Raw scattering intensities for $Q$-scans along the $[H, H, 0]$ direction for various temperatures at $\hbar \omega =0.5$, 1.5, and 4.0 meV.  Peaks in scattering are fitted by Gaussians on a linear background.  A clear gap in scattering at $\hbar \omega =0.5$ meV persists above 40 K until the three-dimensional order breaks down above $T_N$. d-e) Measured dynamic susceptibility at various temperatures for $\hbar \omega =0.5$, 1.5, and 4.0 meV.  Dashed lines are Gaussian fits to $T$=218 K $\chi^{\prime \prime}({\bf{Q}}, 0.5$ meV) in panel (d), $T$=2 K $\chi^{\prime \prime}({\bf{Q}}, 1.5$ meV) in panel (e), and $T$=2 K $\chi^{\prime \prime}({\bf{Q}}, 4.0$ meV) in panel (f).}
\end{figure}

\section{PLCCO, $T_c$=21 K, $T_N$=40 K}

\begin{figure}
\includegraphics[scale=.40]{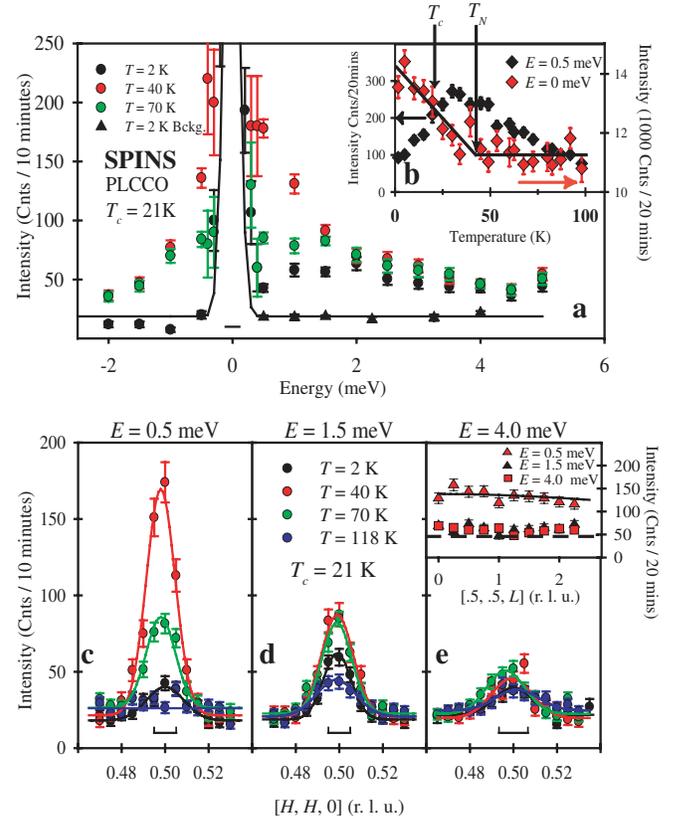}
\caption{
 Temperature dependence of low energy spin excitations measured on the $T_c=21$ K sample.  a)  Representative constant-$Q$ scans at 2 K , 40 K, and 70 K collected at ${\bf Q}=(0.5, 0.5, 0)$ (circles), where the solid black line shows the background scattering collected at ${\bf Q}=(0.53, 0.53, 0)$ (triangles).  At $T = 2$ K, a clear enhancement can be seen around $\hbar\omega=2$ meV.  The solid bracket underneath the incoherent elastic line is the measured energy resolution by a vanadium standard. b)  The temperature dependence of the $\hbar\omega= 0$ meV and $\hbar\omega= 0.5$ meV scattering at $Q = (0.5,0.5,0)$.  The arrows mark the onset of $T_{N}$ and $T_{c}$.  c-e) $Q$-scans of the $T_c=21$ K PLCCO along $[H,H,0]$ for $\hbar\omega= 0.5$, 1.5, and 4 meV and $T = 2$ K, 40 K, 70 K and 118 K.  Center brackets are instrumental resolutions measured by resolution-limited spin-wave peaks from the long-range AF ordered as-grown NSC sample. The inset in panel (e) shows $Q$-scans along the $c$-axis, where solid line is the Cu$^{2+}$ form factor squared and the dashed line marks the background. }
\end{figure}

We now turn to study an underdoped, superconducting PLCCO sample that is annealed to a $T_c=21$ K with a coexisting $T_N\approx 40$ K \cite{dai05}.  The resulting low energy fluctuations around the AF ordering wave vector for this sample are plotted in Fig. 5.  Raw scattering intensities from energy scans at ${\bf Q}=(0.5, 0.5, 0)$ from $-2.0\leq \hbar\omega \leq 5.0$ meV are shown in Fig. 5a for temperatures above and below both $T_c$ and $T_N$ as solid circles, whereas the nonmagnetic background collected at $\bf{Q}$=(0.53, 0.53, 0) is overplotted as solid triangles. One immediate difference appearing between the spectra of this underdoped superconductor and the NSC sample is the appearance of a peak in the low-$T$ dynamic susceptibility along with the absence of a low energy spin-gap.  Figures 5c-e show this gapless peak through $Q$-scans at $\hbar \omega$=0.5, 1.5, and 4 meV, where the $T$=2 K response displays 
a clear enhancement at $\hbar \omega$=1.5 meV.  Solid lines show that Gaussian fits centered at $(0.5,0.5,0)$ on linear backgrounds can well describe the observed peaks in Figs. 5 c-e.  The $T=2$ K excitations exhibit widths much broader than the resolution-limited spin-wave peaks of the NSC system (Figs. 3c and 3d).  Calculating the Fourier transforms of these Gaussian peaks yields the minimum in-plane dynamic correlation lengths of $\xi _{min}=123\pm 21$ \AA\ at $\hbar \omega =0.5$ meV, $\xi _{min}=165\pm 20$ \AA\ at $\hbar \omega =1.5$ meV, and $\xi _{min}=121\pm 27$ \AA\ at $\hbar \omega =4.0$ meV. Using the resolution-limited widths of the NSC PLCCO as reference, 
the true in-plane dynamic spin correlation lengths, $\xi$, are calculated to be $\xi =220\pm 21$ \AA\ at $\hbar \omega =1.5$ meV and $\xi =160\pm 47$ \AA\ at $\hbar \omega =4.0$ meV for the $T_c=21$ K PLCCO.  The substantially broader $Q$-widths in Figs. 5c-e than those of the resolution-limited spin waves in the NSC PLCCO suggest that these excitations cannot arise from the classical spin-wave scattering from the three-dimensional static N$\rm \acute{e}$el ordered phase in the sample \cite{yamada,wilson06}.

Upon warming, the low-$T$ peak in the dynamic susceptibility vanishes and excitations at $\hbar \omega \leq3.5$ meV populate upward until the system is warmed above $T_N$.  Above $T_N$, however, there is a crossover in the magnetic response, and the spectral weight for $\hbar \omega \leq3.5$ meV begins to decrease with increasing temperature.  This crossover is also shown in Fig. 5b in which the intensity of the $\hbar\omega=0.5$ meV excitations and the static $\hbar\omega=0$ meV SDW moment are both plotted as a function of temperature.  With increasing temperature from 2 K, the $\hbar\omega=0.5$ meV fluctuations increase in intensity until the breakdown of the static SDW order near $T_N$.  On further warming, these $\hbar\omega=0.5$ meV fluctuations 
begin to damp until disappearing for temperatures above 118 K (Figs. 5b and 5c). 
These excitations are also found to be quasi-two-dimensional within the CuO$_2$ layers through experiments which orient the crystal in the $[H, H, L]$ scattering plane.  This facilitates $[0.5,0.5,L]$ scans 
in the inset of Fig. 5e,
where the $L$-dependence of the spin excitations at 
$\hbar\omega=0.5$, 1.5, 4 meV is rod-like and simply decays following the Cu$^{2+}$ magnetic form factor (solid line).  This weak correlation perpendicular to the CuO$_2$ planes indicates that 
spin excitations are uncorrelated along the $c$-axis and reflect 
the highly anisotropic exchange coupling between Cu sites within the CuO$_2$ plane ($J>100$ meV) 
and the much weaker out-of-plane exchange  \cite{wilson06}.

\begin{figure}
\includegraphics[scale=.40]{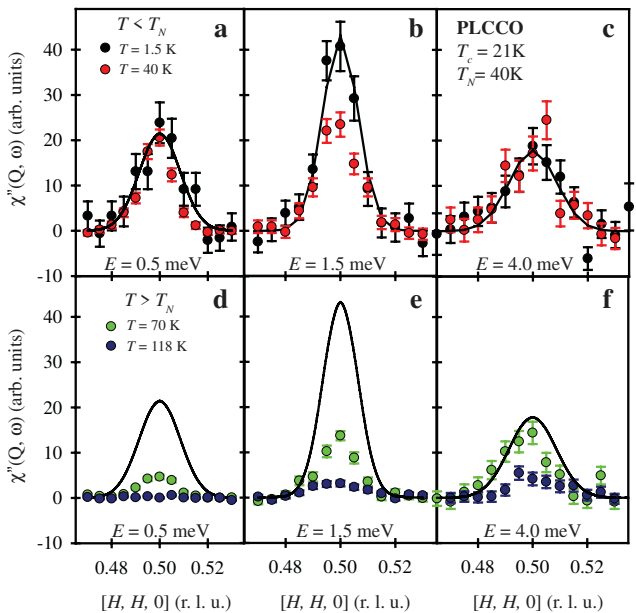}
\caption{
 Temperature dependence of $\chi^{\prime \prime}({\bf{Q}}, \omega)$ at the ${\bf Q}=(0.5, 0.5, 0)$ position for $\hbar \omega=0.5$, 1.5, and 4.0 meV.  a-c) $\chi^{\prime \prime}({\bf{Q}}, \omega)$ for $T\leq T_N$ ($T=2$ K and 40 K).  d-f)  $\chi^{\prime \prime}({\bf{Q}}, \omega)$ for $T\geq T_N$ ($T=70$ K and 118 K).  In all panels, solid lines are Gaussian fits to the $T=2$ K dynamic susceptibility at the corresponding energy.  
 }
\end{figure}
	
For $\hbar \omega >3.5$ meV, the magnetic scattering intensity at ${\bf Q}=(0.5,0.5,0)$ is $T$-independent 
from $T=2$ K to 120 K as shown in Figs. 5a and 5e. In contrast, the spin scattering at $\hbar \omega\leq 3.5$ meV is strongly coupled to the appearance of the AF phase. Hence, this regime of $T$-independent $S({\bf Q}, \omega)$ provides additional evidence of a crossover in the spin dynamics that is now instead reflected in the energy scale of the observed spin fluctuation spectrum.  A changeover in the response of the system can also be tested by plotting $\chi ^{\prime\prime}({\bf Q}, \omega)$ at various energies for temperatures both above and below $T_N$ as shown in Fig. 6. The top panels in Figs. 6a-c show the measured dynamic susceptibility below $T_N$.  From these, the population of the lowest energy excitations at $\hbar \omega =0.5$ meV can be seen to simply follow Bose statistics with $\chi ^{\prime\prime}($$\bf{Q}$$, \omega$) remaining constant.  At $\hbar \omega =1.5$ meV, there exists a slight decrease in the susceptibility upon warming to 40 K, most likely due to the vanishing peak in the susceptibility (which is present at 2 K, see Fig. 5).
The susceptibility at $\hbar \omega =4.0$ meV, however, recovers this behavior of following the Bose population factor with no $T$-dependence in $\chi ^{\prime\prime}({\bf Q}, \omega)$ up to $T=40$ K.  For temperatures above $T_N$, the dynamic susceptibility decreases sharply with increasing $T$ for all energies studied as shown in Figs. 6d-f.  This is simply reflective of the decrease in magnetic scattering with increasing $T$ (for $T>T_N$) at $\hbar\omega =0.5$ and 1.5 meV.  The decrease in $\chi ^{\prime\prime}({\bf Q},\omega)$ at $\hbar \omega =4.0$ meV, instead, arises from the observed $T$-independent $S({\bf Q}, \omega)$ for temperatures up to 120 K.
This is in sharp constrast to the $\hbar\omega =4.0$ meV spin-wave excitations in as-grown PLCCO, where
the intensity of $S({\bf Q}, \omega)$ is entirely controlled by the Bose statistics (Figs. 4c and 4f).
An interesting question then arises:  How do these regimes of 
energy and temperature $(\hbar \omega, T)$ that couple to either the AF phase or the paramagnetic phase (giving rise to this $T$-independent response in $S({\bf Q},\omega)$) evolve with increased doping?  We address this question below through both the experimental observation (Sections V and VI) 
and data analysis (Section VII).

\section{PLCCO, $T_c$=23 K, $T_N$=25 K}

We now describe results on an PLCCO sample with $T_c=23$ K (Fig. 1b). 
Experiments probing the static magnetic ordering of this system 
show an AF Bragg reflection at the ${\bf Q}=(0.5,0.5,1)$ position with a $T_N\approx$ 25 K (Fig. 2c).  
$Q$-scans through this AF ordering wave vector along the
$[0.5,0.5,L]$ direction
 reveal that this weak reflection disappears for $T>25$ K (Fig. 2d).  In Fig. 7a, the measured scattering at ${\bf Q}=(0.5, 0.5, 0)$ for $-1.5\leq \hbar \omega \leq 5.0$ meV is plotted as crossed-boxes, while nonmagnetic background collected at ${\bf Q}=(0.56, 0.56, 0)$ is plotted as solid triangles. Examining these low-$E$ fluctuations, substantial differences appear between the spectra of this $T_c=23$ K sample and those of the $T_c=21$ K system.  The excitations at the ${\bf Q}=(0.5, 0.5, 0)$ position remain gapless; however, they lack a clearly defined peak in the low-$T$ susceptibility at $T=2$ K.  A comparison of the $T=2$ K magnetic excitations at $\hbar \omega=0.5$, 1.5, and 4.0 meV is shown between the $T_c=23$ K and $T_c=21$ K PLCCO samples in Figs. 7 c-e.  The dashed lines show Gaussian fits to the scattering observed in the $T_c=21$ K sample whose enhancement at $\hbar \omega =1.5$ meV contrasts the continual decrease of scattering intensity with increasing energy transfer seen in the $T_c=23$ K PLCCO (crossed-box symbols).  Additionally, these plots show that the excitations observed in the $T_c=23$ K system have broadened in $Q$ at all energies relative those observed in the $T_c=21$ K sample.  The Fourier transforms of the Gaussian fits in Figs. 7c-e give minimum in-plane dynamic spin correlation lengths of $\xi _{min}=86\pm 10$ \AA\ at $\hbar \omega =0.5$ meV, $\xi _{min}=87\pm 12$ \AA\ at $\hbar \omega =1.5$ meV, and $\xi _{min}=67\pm 11$ \AA\ at $\hbar \omega =4.0$ meV.  Correcting for the 
 instrumental resolution yields $\xi =93\pm 14$ \AA\ at $\hbar \omega =1.5$ meV and $\xi =72\pm 13$ \AA\ at $\hbar \omega =4.0$ meV for the $T_c=23$ K PLCCO.

\begin{figure}
\includegraphics[scale=.40]{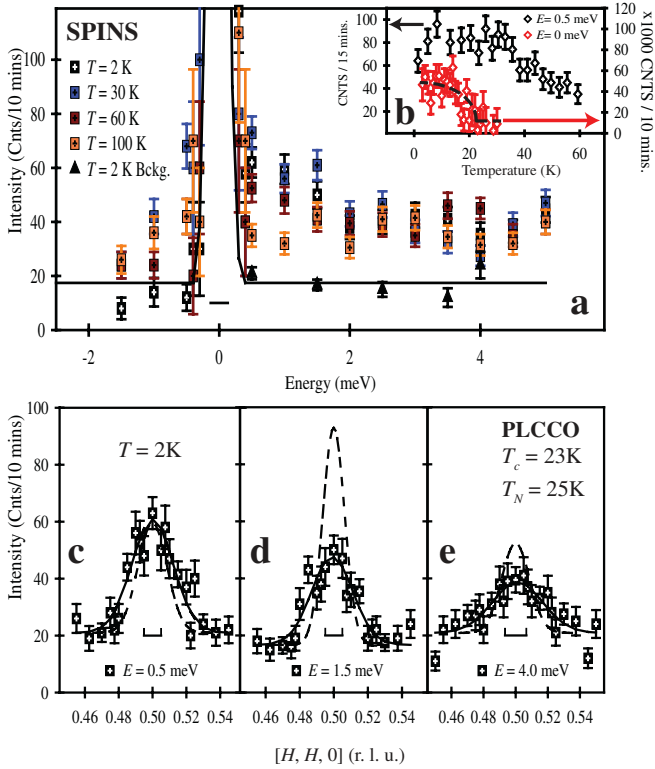}
\caption{
Temperature dependence of the low energy 
spin fluctuations in the $T_c=23$ K PLCCO.  a) Constant-$Q$ scans at ${\bf Q}=(0.5,0.5,0)$ and  
2 K, 30 K, 60 K and 100 K. The solid black line shows the background scattering at ${\bf Q}=(0.56,0.56,0)$ 
and the horizontal bar beneath the incoherent elastic peak is the instrumental resolution.  b) Temperature dependence of the inelastic ($\hbar\omega=0.5$ meV) scattering at ${\bf Q}=(0.5,0.5,0)$ and elastic ($\hbar\omega=0$ meV) scattering at ${\bf Q}=(0.5,0.5,1)$ Bragg position also shown in Fig. 2c demonstrating a magnetic order at $T_N\approx 25$ K.  c-e) $Q$-scans of the $T_c=23$ K PLCCO along the $[H,H,0]$ direction 
for $\hbar\omega= 0.5$, 1.5, and 4 meV, respectively, at $T = 2$ K.  The dashed lines show identical scans from the $T_c=21$ K sample with its $(1,1,0)$ Bragg intensity normalized to that of the $T_c=23$ K sample.  Center brackets are instrumental resolutions and solid lines are Gaussian fits on flat backgrounds.}
\end{figure}

The temperature dependence of the magnetic excitations in Fig. 7 a shows a similar type of crossover in the spin dynamics to those observed in the under-doped $T_c=21$ K system.  The regimes of these two types of spin dynamics are again determined by the energy and temperature scale of the AF order in the system, now with a $T_N\approx$ 25 K.  For $T<T_N$, there exists a slight enhancement upon warming of the spin fluctuations at $\hbar \omega <2.5$ meV, whereas for $T>T_N$ and $\hbar \omega \geq 2.5$ meV the observed scattering intensity remains $T$-independent over a broad temperature range ($T=2$ K $\rightarrow$ 60 K).  The relative intensity of the magnetic scattering at $T=100$ K is difficult to determine due to uncertainties in the nonmagnetic background contributions, whereas there was no observed change in background scattering between $T=2$ K and 60 K. The enhancement coupled to the AF order is significantly damped and no longer strictly follows the Bose population factor.  Instead, a weak increase in the population of the $\hbar\omega=0.5$ meV fluctuations is observed (as shown in Fig. 7b) as the system is warmed towards $T_N$.  The coupling of these low energy fluctuations to $T_N$ as a function of $T$ is also significantly broadened (Fig. 7b), in contrast to that of the $T_c=21$ K PLCCO (Fig. 5b). 
On the other hand, the temperature and energy region in which 
the magnetic scattering intensity is temperature independent increases, reflecting 
the much weaker AF phase in this sample along with the reduced energy scale of $T_N$.  

\begin{figure}
\includegraphics[scale=.40]{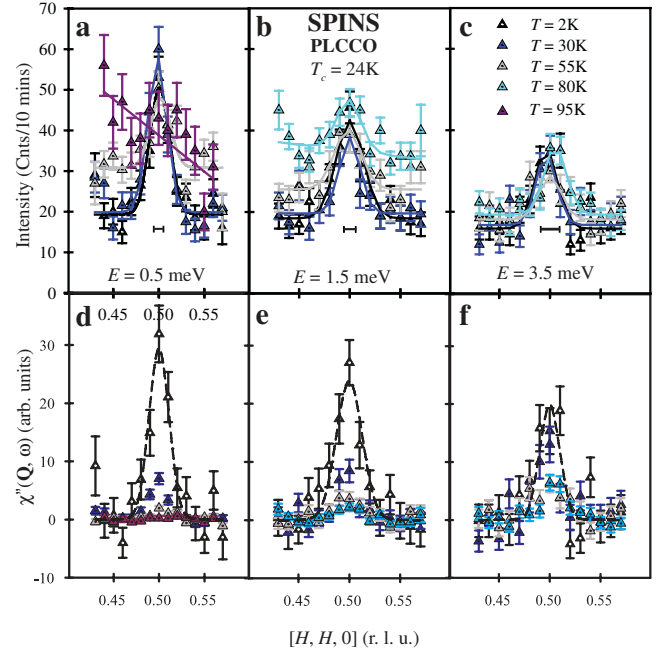}
\caption{
Temperature dependence of the low energy spin fluctuations and dynamic susceptibility in the $T_c=24$ K sample.  a-c) Constant-$E$ scans at $T=2$, 30, 55, 80, and 95 K through the ${\bf Q}=(0.5, 0.5, 0)$ position along
the $[H, H, 0]$ direction.  Solid lines are Gaussian fits on linear backgrounds.  The $T=2$, 30, and 55 K data are replotted for comparison from Ref. 11.  Solid brackets show the resolution of the spectrometer. 
The increased background scattering with increasing temperature 
at $\hbar\omega=0.5$ and 1.5 meV may arise from single/multi-phonon scattering
and/or air scattering. d-f) $\chi^{\prime \prime}({\bf{Q}}, \omega)$ at $T=2$, 30, 55, 80, and 95 K.  Dashed lines are Gaussian fits to the $T=2$ K dynamic susceptibility.}
\end{figure}

\section{PLCCO, $T_c=24$ K, $T_N<600$ ${\rm m}$K}  

Turning now to the final optimally doped $T_c=24$ K sample, previous experiments have shown that there exists no static AF order co-existing with superconductivity in this sample down to 
600 mK \cite{wilson}.  The low energy excitations for this system were first reported in Ref. 11 and are expanded upon in Fig. 8.  Figures 8a-c show the raw scattering intensities observed for $\hbar \omega =0.5$, 1.5, and 3.5 meV at various temperatures.  The spin excitations in these energies are fitted by Gaussians on linear backgrounds and give minimum in-plane dynamic spin correlation lengths of 
$\xi _{min}=96\pm 15$ \AA\ at $\hbar \omega =0.5$ meV, $\xi _{min}=80\pm 10$ \AA\ at $\hbar\omega =1.5$ meV, and $\xi _{min}=94\pm 24$ \AA\ at $\hbar \omega =3.5$ meV at $T=2$ K \cite{wilson}.  These widths are within error to those observed in the $T_c=23$ K sample when neglecting the slight change in instrumental resolution on changing $E_f$ from 5.0 meV to 3.7 meV.  This is reasonable since the measured $Q$-widths are appreciably larger than the instrument resolution in both geometries \cite{wilson}.

Upon warming from $T=2$ K to $T=30$ K, there is no change in the measured magnetic scattering intensity at all energies down to $\hbar \omega =0.5$ meV.  Further increase in temperature to $T=55$ K renders a slight reduction in scattering at $\hbar\omega=0.5$ meV and 1.5 meV, with no change measured in the 3.5 meV excitations.  At $T=95$ K, the peak at $\hbar \omega =0.5$ meV has completely vanished.  There exists a strong suppression of the $\hbar\omega=1.5$ meV fluctuations at $T=80$ K.  In contrast, there are no changes in the $\hbar\omega=3.5$ meV fluctuations up to 80 K.  Figures 8d-e show the measured $\chi ^{\prime \prime}({\bf Q},\omega )$ for the same energies reflecting a continued decrease in the susceptibility with increasing temperature.  The dashed lines show Gaussian fits centered at $(0.5, 0.5, 0)$ for the $T=2$ K susceptibility to highlight 
the dramatic decrease in $\chi ^{\prime \prime}({\bf Q},\omega )$ at different energies (Figs. 8d-f).
At the highest measured temperature ($T=95$ K), the system becomes gapped at $\hbar \omega =0.5$ meV 
similar to the two other underdoped PLCCO.  The absence of any regime in which the dynamic susceptibility remains constant, reflective of bosonic excitations similar to those observed in the $T_c=21$ K samples, demonstrates a drastic deviation from the two-distinct regimes of magnetic response observed in the 
$T_c=21$ K PLCCO.  Therefore, low energy spin excitations in PLCCO evolve from coupling 
to the onset of the AF phase in underdoped materials to essentially temperature independent from 2 K
to 30 K for the optimally doped sample.

Recently, we have discovered that optimally doped PLCCO has a resonance \cite{wilson}.  Similar to hole-doped materials \cite{dai01}, the resonance in electron-doped PLCCO increases in intensity below $T_c$ and its energy scales with $5.8k_BT_c$ forming a universal plot for all superconducting copper oxides irrespective of electron- or hole-doping \cite{wilson}.  However, in contrast to hole-doped materials, magnetic excitations below the resonance in electron-doped PLCCO form commensurate and gapless scattering (Fig. 8). Therefore, the hour-glass shaped dispersion in the magnetic scattering of hole-doped superconducting materials \cite{woo,hayden04,tranquada04,christen04,tranquada05} may not be a universal feature of all high-$T_c$ superconductors.  Instead, the resonance itself appears to be a universal property of the superconducting copper oxides.
                          
\section{Discussion} 

\begin{figure}
\includegraphics[scale=.40]{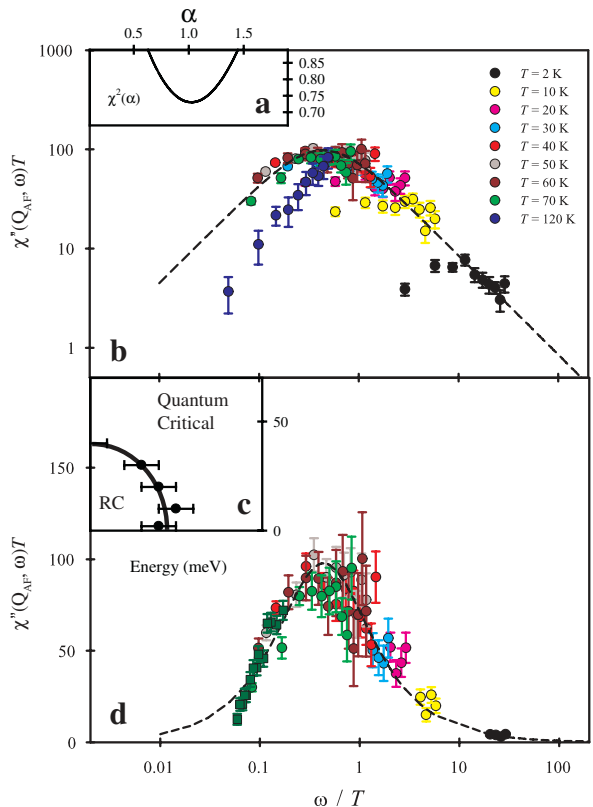}
\caption{
$\omega/T$ scaling in the dynamic susceptibility of PLCCO with $T_c=21$ K and $T_N=40$ K.
a)  The results from a minimization of $\chi^2(\alpha)$ for all $\chi^{\prime\prime}(\omega,Q)$ data with 80 K $\ge T \ge$ 30 K and for data with $T<30$ K and $\hbar\omega> 2$ meV.   The bin size of 0.5 along the log($\omega/T)$ scale is the only assumption made in obtaining $\alpha = 1$. b)  The log plot of $\chi^{\prime\prime}T$ as a function of $\omega/T$ shows three distinct regions:  1) data with 80 K $\ge T \ge$ 30 K and with $T<$ 30 K and $\hbar\omega> 2$ meV are within the quantum critical scaling regime and collapse onto a universal curve. 2) the data at 118 K show a clear high temperature departure from this quantum critical scaling behavior. 3) the data with  $T<30$ K and $\hbar\omega<2$ meV show a low temperature departure from the scaling regime.   The breakdown of scaling at 120 K could be due to uncertainties in determining the backgrounds. c)  Summary of the $\omega/T$ scaling regime. d) $\chi^{\prime\prime}T$ as a function of $\omega/T$ now plotted with only the data in the quantum critical scaling regime shown in panel (c).  This provides a clearer picture of the universal collapse for data within the valid scaling regime. 
 }
\end{figure}

The identification of the continuous suppression of $T_{N}$ as the optimal superconductivity is approached in PLCCO as a function of annealing process \cite{dai05} suggests the possibility of a magnetic QCP, regardless of the precise nature of the magnetic structure (homogeneous or inhomogeneous).  To fully establish the existence of such a QCP requires the study of magnetic dynamics, which we have shown above using inelastic neutron scattering.  We focus on two signatures of a QCP.  First, a QCP is accompanied \cite{hertz,chakravarty,sachdev2,millis,varma,abanov,si,sachdev3} by a quantum critical region at finite temperatures and finite energies -- bounded below by a scale which gradually goes to zero as the QCP is reached -- where the dynamics manifests the excitations of the QCP.  Second, in this quantum critical regime, the dynamics are scale-invariant, a particular form of which is an $\omega/T$ scaling \cite{sachdev2,varma};  such an $\omega/T$ scaling has been systematically studied in heavy Fermion metals \cite{schroeder,aronson,wilson05} in which the existence of a magnetic QCP is not in doubt \cite{stewart}.  In the following, 
we will determine how our data can be described by the standard QCP theory.

\begin{figure}
\includegraphics[keepaspectratio=true,width=0.95\columnwidth,clip]{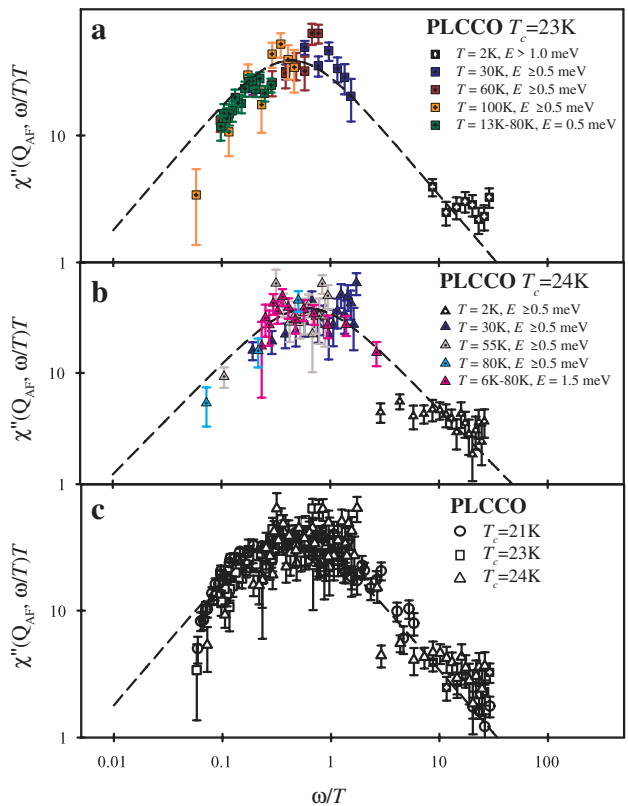}
\caption{
$\omega/T$ scaling in the dynamic susceptibility of PLCCO with $T_c=23$ K, $T_N=25$ K and $T_c=24$ K (with no coexisting AF order).  a) A log plot of $\chi^{\prime\prime}T$ as a function of $\omega/T$ for the $T_c=23$ K sample over-plotted on the universal curve from Fig. 9b. By normalizing the $(1,1,0)$ Bragg intensities of both samples, we find $A=4.5$ which indicates that magnetic scattering at ${\bf Q}=(0.5,0.5,0)$ is 
considerably weaker in the 
$T_c=23$ K PLCCO.  This scaling plot reflects the modified scaling regimes in this $T_c=23$ K sample as discussed in the text.  b) $\chi^{\prime\prime}T$ plotted as a function of $\omega/T$ for the $T_c=24$ K sample again over-plotted on the universal curve from Fig. 9b. c) Scaling relation now plotted for all three superconducting samples, $T_c=21$ K, $T_c=23$ K, and $T_c=24$ K, and their respective scaling regimes showing a universal collapse onto the curve from Fig. 9b.  
 }
\end{figure} 

To perform a scaling analysis for the $T_c=21$ K PLCCO, we note that the $\omega/T$ scaling can be written as $\chi^{\prime\prime}(Q,\omega)T^\alpha = F(Q, \omega/T)$, where the scaling exponent $\alpha$ and the scaling function $F(Q,\omega/T)$ are determined through the best-observed collapse of the data onto 
one universal curve \cite{bao}. The quantum critical scaling is different from the critical scattering from classical AF second order phase transition as the former is controlled by $T$ itself while the characteristic energy scale $\omega_{c}$ in the latter case is determined by reduced temperature $t=|T-T_N|/T_N$ ($|t|\ll 1$) \cite{collin}.  For the Heisenberg antiferromagnet with $T_N=40$ K, $\omega_{c}$ should fall within the quasi-elastic part of the energy spectra and thus not contribute to the observed magnetic scattering in Fig. 5. By plotting all data above the 30 K energy scale ($\hbar\omega\geq 3$ meV) as a function of $\omega/T$ and minimizing $\chi^{2}$ (Ref. 38), we find $\alpha=1$ independent of the functional form of $F(Q,\omega/T)$ (Fig. 9a).  Using this value for the exponent leads to a scaling plot of $\chi^{\prime\prime}(Q,\omega)T$ as a function of $\omega/T$ (Fig. 9b).  Such a plot also identifies the scaling regime, which is bounded at low energy (2 meV) and temperatures (Fig. 9c).  The data within the scaling regime in Fig. 9c clearly show a collapse onto a universal curve as plotted in Fig. 9d.  We fit the latter using $F(y)\propto{y}/(1+(y/C)^2)$ and find $C= 0.44\pm 0.02$.  This collapse of the data over more than two decades of $\omega/T$ strongly suggests the presence of universal dynamics.  If these universal dynamics indeed reflect a nearby QCP, the scaling behavior must break down at low energies and temperatures, with the cut-off scale determined by the distance to the QCP.  Indeed, for small $\omega$ the $\omega/T$ scaling is observed only at temperatures above about 30 K (Fig. 9b).  The fact that this temperature scale is of the order of $T_{N}$ suggests that the cut-off to scaling is connected with the development of the AF order.  Likewise, at the lowest measured temperature, $T = 2$ K, the $\omega/T$ scaling is seen only at $\hbar\omega$ above about 2 meV (Fig. 9b), which, within the error bars, is equal to the low-$\omega$ cut-off temperature multiplied by the universal constant $C$. 

Now turning to how this scaling regime evolves in samples tuned closer to optimal doping, the same type of scaling analysis can be performed for the low-energy magnetic spectra of the $T_c=23$ K PLCCO.  Since our neutron diffraction measurements at ${\bf Q}=(0.5,0.5,1)$ show a $T_N\approx25$ K (Fig. 2c), this sample should be much closer to the magnetic QCP than the $T_c=21$ K ($T_N=40$ K) PLCCO.  Figure 10a shows the summary of the $\chi^{\prime\prime}(Q,\omega)T$ of the $T_c=23$ K PLCCO over-plotted on the universal scaling fit to the $T_c=21$ K sample (dashed line).  A scale factor, $A$, was used to normalize the $\chi^{\prime\prime}(Q,\omega)$ at ${\bf Q}=(0.5,0.5,0)$ for the $T_c=23$ K PLCCO to that of the $T_c=21$ K PLCCO, which reflects the fact that the $Q$-width in the former is much broader (Figs. 7c-e).  Besides the overall scale factor, we find that all data with $\hbar\omega\ge 1$ meV and/or $T\ge 25$ K  fall on the universal curve, consistent with the $T_c=23$ K PLCCO being closer to the QCP point. The $\omega/T$ scaling regime is seemingly correlated with the gradual suppression of the AF order as optimal superconductivity is approached, indicating the presence of a magnetic quantum critical point in electron-doped superconductors.  
However, the dynamic in-plane spin correlation length as stated earlier is observed to decrease with increasing $T_c$ and decreasing $T_N$, suggesting that this QCP in PLCCO cannot arise from criticality towards a three-dimensional  
long-range AF order at 0 K \cite{sachdev2}.  Instead, the data are similar to what happens in the Li-doped La$_2$CuO$_4$ \cite{bao} and in certain spin-glass QCP heavy Fermion systems \cite{aronson,wilson05}. Although there is at present no comprehensive theory for the expected properties of a metallic spin-glass QCP, the reduced magnetic transition temperature and the continuous nature of the transition (Fig. 2c) imply that the scaling behavior in the $T_c=23$ K PLCCO should persist down to lower energies and temperatures than that of the $T_c=21$ K sample.

Continuing this analysis on the magnetic fluctuations observed in the $T_c=24$ K sample, an identical scaling plot is shown in Fig. 10b.  A constant scale factor was again used to normalize the scaling fit from the $T_c=21$ K sample to overlay with the data from the $T_c=24$ K sample.  As the system is tuned closer to the QCP, there should now be larger ranges of energies and temperatures $(\hbar \omega, T)$ in which the scaling remains valid.  Indeed, now all energies and temperatures (for which a suitable nonmagnetic background was measured) are shown to collapse on the same universal curve determined for the $T_c=21$ K sample (dashed line in Fig. 10b).  This encompasses the entire range of energy and temperature probed from $0.5\geq \hbar\omega \geq 4.0$ meV and $2\geq T\geq 80$ K with the exception of a notable divergence at the $\hbar\omega=0.5$ meV spin fluctuations measured at $T=2$ K in Fig. 10b.  
This divergence from the scaling fit for spin dynamics with $\hbar\omega\leq 1.25$ meV may result from the system being tuned slightly beyond the QCP.  In this case, the low energy/temperature spin dynamics crossover into the quantum disordered regime and no longer obey the scaling relation of quatum critical excitations.  
Since the system does not exhibit AF order to at least 600 mK, we cannot determine how close
the sample is to the $T_N=0$ K QCP. 
Nevertheless, the overall trend for the three superconducting PLCCO investigated 
is that of increasing regimes of validity for the dynamic $\omega/T$ scaling as 
the system is tuned closer to the QCP itself.  

Finally, the valid scaling regimes for all three SC PLCCO samples are overplotted in Fig. 10c showing a universal collapse onto a common function, with the exception of the lowest energy excitations at $T=2$ K for the $T_c=24$ K sample as discussed earlier.  The universal collapse signifying $\omega/T$ scaling 
and the systematic evolution of scaling regime as AF order is suppressed with annealing strongly
suggest the presence of a magnetic QCP in the electron-doped PLCCO.  
Recent transport and optical measurements on electron-doped Pr$_{2-x}$Ce$_{x}$CuO$_{4}$ (PCCO)
have identified singular behaviors, which appear compatible with the influence of a QCP \cite{dagan,zimmers}.  
Our results suggest that such a QCP may have a magnetic origin, possibly due to the presence of a 
spin-glass QCP.  We note that previous work has shown the presence of a 
spin-glass QCP in the hole-doped cuprates \cite{Panagopoulos}, thus suggesting that this magnetic QCP might be
a common feature in high-$T_c$ superconductors regardless of doped-carrier type.

\section{Conclusions} 

We have systematically measured the doping evolution of the low energy spin fluctuations in the electron-doped cuprate, PLCCO, as the system is tuned from an as-grown NSC antiferromagnet  
into a phase-pure optimally doped superconductor.  The as-grown, semiconducting PLCCO system exhibits gapped low energy spin fluctuations consistent with those observed in the parent compound Pr$_2$CuO$_4$.  Tuning the system into an optimally doped superconductor changes these low energy spin excitations dramatically.  
Instead of a gapped spin wave spectrum, the superconducting PLCCO samples exhibit gapless low energy spin dynamics which exhibit two differing regimes of response in $(\hbar \omega, T)$: the first of these coupling to the onset of the AF order in the system and the second appearing as spin dynamics whose scattering is weakly temperature dependent over large temperature ranges.  The regime that couples to the AF order in these samples is seen to evolve from $T<40$ K and $\hbar\omega <2.5$ meV in the $T_c=21$ K sample to $T<25$ K and $\hbar\omega<1.5$ meV in the $T_c=23$ K PLCCO.  At optimal doping ($T_c=24$ K), where the static AF phase is suppressed to below
600 mK, the observed magnetic scattering for $T\leq 30$ K or $\hbar\omega\geq 1.5$ meV is $T$-independent, suggestive of the possible influence of quantum critical fluctuations in the system.  Subsequent scaling analysis of the data revealed a collapse of appropriate regions of $(\hbar \omega, T)$ onto a universal curve for all three superconducting PLCCO systems studied, thereby providing microscopic evidence for a QCP in the electron doped PLCCO.  Additionally, the first system discussed, $T_c=21$ K PLCCO, displays a peak in the low-$T$ susceptibility centered at $\Gamma _{0}\approx 2.0$ meV which vanishes upon warming, whereas systems tuned closer to optimal doping ($T_c=23$ K and $T_c=24$ K) display only continuously decreasing magnetic response with increasing energy transfer.  Our experiments described here have provided
a systematic investigation into the evolution of the low energy spin dynamics in an electron-doped 
copper oxide, PLCCO, thereby providing valuable constraints on microscopic theories seeking to model the spin excitations as the materials evolve from a long-range ordered antiferromagnet into an optimally doped superconductor.

\section{Acknowledgements}
We are grateful to Hong Ding, J. W. Lynn, H. A. Mook, 
Tai Kai Ng, and F. C. Zhang for helpful discussions.
The neutron scattering part of this work is supported by the U.S. NSF DMR-0453804.  The PLCCO 
sample growth is supported
in part by U.S. DOE DE-FG02-05ER46202. ORNL is supported by 
DE-AC05-00OR22725 with UT/Battelle LLC. LANL is supported by U.S. DOE.
SPINS is supported by the U.S. NSF through Grant No. DMR-9986442. 
The part of the work done in Japan was supported by the
Grant-in-Aid for Science provided by the Japan Society for
the Promotion of Science. Q.S. is supported by NSF DMR-0424125 and
the Robert A. Welch foundation.


\begin{thebibliography}{}

\bibitem{hayden96} S. M. Hayden, G. Aeppli, H. A. Mook, T. G. Perring, T. E. Mason, S.-W. Cheong,
and Z. Fisk, Phys. Rev. Lett. {\bf 76}, 1344 (1996).
\bibitem{bourges97} P. Bourges, H. Casalta, A. S. Ivanov, and D. Petitgrand, 
Phys. Rev. Lett. {\bf 79}, 4906 (1997).
\bibitem{coldea} R. Coldea,S. M. Hayden, G. Aeppli, T. G. Perring, C. D. Frost, T. E. Mason,
S.-W Cheong, and Z. Fisk, Phys. Rev. Lett. {\bf 86}, 5377 (2001).
\bibitem{scalapino} D. J. Scalapino, Science {\bf 284}, 1282 (1999).
\bibitem{dai99} Pengcheng Dai, H. A. Mook, S. M. Hayden, G. Aeppli, T. G. Perring, R. D. Hunt, and F. Do$\rm\breve{g}$an, Science {\bf 284} 1344 (1999).
\bibitem{woo} Hyungje Woo, Pengcheng Dai, S. M. Hayden, H. A. Mook, T. Dahm, 
D. J. Scalapino, T. G. Perring, F. Do$\rm\breve{g}$an, cond-mat/0608280 (Nat. Phys. in press).
\bibitem{hayden04} S. M. Hayden, H. A. Mook, Pengcheng Dai, T. G. Perring, and F. Do$\rm\breve{g}$an, Nature (London) {\bf 429}, 531 (2004).
\bibitem{tranquada04} J. M. Tranquada, H. Woo, T. G. Perring, H. Goka, G. D. Cu, G. Xu,
M. Fujita, and K. Yamada, Nature (London) {\bf 429}, 534 (2004).
\bibitem{christen04} N. B. Christensen, D. F. McMorrow, H. M. Ronnow, B. Lake, S. M. Hayden, G. Aeppli,
T. G. Perring, M. Mangkorntong, M. Nohara, and H. Takagi, Phys. Rev. Lett. {\bf 93}, 147002 (2004).
\bibitem{tranquada05} John M. Tranquada , cond-mat/0512115 (2005).
\bibitem{yamada} K. Yamada, K. Kurahashi, T. Uefuji, M. Fujita, S. Park, S.-H. Lee, and
Y. Endoh, Phys. Rev. Lett. {\bf 90}, 137004 (2003).
\bibitem{wilson} Stephen D. Wilson, Pengcheng Dai, Shiliang Li, Songxue Chi, H. J. Kang,
and J. W. Lynn, Nature (London) {\bf 442}, 59 (2006).
\bibitem{wilson06} Stephen D. Wilson, Shiliang Li, Hyungje Woo, Pengcheng Dai, H. A. Mook,
C. D. Frost, Seiki Komiya, and Yoichi Ando, Phys. Rev. Lett. \textbf{96}, 157001 (2006).
\bibitem{yamada2} K. Yamada, C. H. Lee, K. Kurahashi, J. Wada, S. Wakimoto, S. Ueki,
H. Kimura, Y. Endoh, S. Hosoya, G. Shirane, R. J. Birgeneau, M. Greven, M. A. Kastner,
and Y. J. Kim, Phys. Rev. B {\bf 57}, 6165 (1998).
\bibitem{dai01} Pengcheng Dai, H. A. Mook, R. D. Hunt, and F. Do$\rm\breve{g}$an, 
Phys. Rev. B {\bf 63}, 054525 (2001).
\bibitem{boothroyd} A. T. Boothroyd, S. M. Doyle, D. McK. Paul, and  R. Osborn, Phys. Rev. B {\bf 45}, 10075 (1992).
\bibitem{sumarlin} I. W. Sumarlin, J. W. Lynn, T. Chattopadhyay, S. N. Barilo, D. I. Zhigunov, 
and J. L. Peng, Phys. Rev. B {\bf 51}, 5824 (1995).
\bibitem{fujita} M. Fujita, T. Kubo, S. Kuroshima, T. Uefuji, K. Kawashima, K. Yamada,
I Watanabe, and K. Nagamine, Phys. Rev. B {\bf 67}, 014514 (2003).
\bibitem{dai05} Pengcheng Dai, H. J. Kang, H. A. Mook, M. Matsuura, J. W. Lynn, Y. Kurita,
S. Komiya, and Y. Ando, Phys. Rev. B {\bf 71}, 100502(R) (2005).
\bibitem{kang05} H. J. Kang, Pengcheng Dai, H. A. Mook, D. N. Argyriou, V. Sikolenko, J. W. Lynn,
Y. Kurita, S. Komiya, and Y. Ando, Phys. Rev. B {\bf 71}, 214512 (2005).
\bibitem{uefuji} T. Uefuji,T. Kubo, K. Yamada, M. Fujita, K. Kurahashi, I. Watanabe,
and K. Nagamine, Physica C {\bf 357-360}, 208 (2001).
\bibitem{kang03} H. J. Kang, Pengcheng Dai, J. W. Lynn, M. Matsuura, J. R. Thompson, Shou Cheng Zhang,
D. N. Argyriou, Y. Onose, and Y. Tokura, Nature (London) {\bf 423}, 522 (2003).
\bibitem{matsuura03} M. Matsuura, Pengcheng Dai, H. J. Kang, J. W. Lynn, D. N. Argyriou,
K. Prokes, Y. Onose, and Y. Tokura, Phys. Rev. B {\bf 68}, 144503 (2003).
\bibitem{tokura} Y. Tokura, H. Takagi, and S. Uchida, Nature (London) {\bf 337}, 345 (1989).
\bibitem{schroeder} A. Schroder, G. Aeppli, R. Coldea, M. Adams, O. Stockert,
H. v. Lohneysen, E. Bucher, R. Ramazashvili, P. Coleman, 
Nature (London) \textbf{407}, 351 (2000).
\bibitem{aronson} M. C. Aronson, R. Osborn, R. A. Robinson, J. W. Lynn,
R. Chau, C. L. Seaman, and M. B. Maple, Phys. Rev. Lett. \textbf{87}, 197205 (2001).
\bibitem{wilson05} Stephen D. Wilson, Pengcheng Dai, D. T. Adroja, S.-H. Lee, J.-H. Chung, J. W. Lynn, N. P.
Butch, and M. B. Maple, Phys. Rev. Lett. {\bf 94}, 056402 (2005).
\bibitem{stewart} G. R. Stewart, Rev. Mod. Phys. \textbf{73}, 797 (2001).
\bibitem{lavrov04} A. N. Lavrov, H. J. Kang, Y. Kurita, T. Suzuki, S. Komiya, J. W. Lynn,
S.-H. Lee, Pengcheng Dai, and Y. Ando, Phys. Rev. Lett. {\bf 92}, 227003 (2004).
\bibitem{petrigrand} D. Petitgrand, S. V. Maleyev, Ph. Bourges, and A. S. Ivanov, 
Phys. Rev. B {\bf 59}, 1079 (1999).
\bibitem{hertz} J. A. Hertz, Phys. Rev. B {\bf 14}, 1165 (1976).
\bibitem{chakravarty} Sudip Chakravarty, B. I. Halperin, and D. R. Nelson, 
Phys. Rev. B. {\bf 39}, 2344 (1989).
\bibitem{sachdev2} S. Sachdev and J. W. Ye, Phys. Rev. Lett. {\bf 69}, 2411 (1992).
\bibitem{millis} A. J. Millis, Phys. Rev. B {\bf 48}, 7183 (1993).
\bibitem{varma} C. M. Varma, Phys. Rev. Lett. {\bf 83}, 3538 (1999).
\bibitem{abanov} A. Abanov and A. Chubukov, Phys. Rev. Lett. {\bf 84}, 5608 (2000).
\bibitem{si} Qimiao Si, S. Robello, K. Ingersent, J. Lleweilun Smith, Nature (London) {\bf 413}, 804 (2001).
\bibitem{sachdev3} S. Sachdev, Rev. Mod. Phys. {\bf 75}, 913 (2003).
\bibitem{bao} W. Bao, Y. Chen, Y. Qiu, and J. L. Sarrao, Phys. Rev. Lett. \textbf{91}, 127005 (2003).
\bibitem{collin} M. F. Collins, Magnetic Critical Scattering (Oxford University Press, Oxford, 1989) page 46.
\bibitem{dagan} Y. Dagan, M. M. Qazilbash, C. P. Hill, V. N. Kulkarni, and R. L. Greene, 
Phys. Rev. Lett. \textbf{92}, 167001 (2004).
\bibitem{zimmers} A. Zimmers, J. M. Tomczak, R. P. S. M. Lobo, N. Bontemps, C. P. Hill,
M. C. Barr, Y. Dagan, R. L. Greene, A. J. Millis, and C. C. Homes, Europhys. Lett. {\bf 70}, 225 (2005).
\bibitem{Panagopoulos} C. Panagopoulos, J. L. Tallon, B. D. Rainford, T. Xiang, J. R. Cooper, and C. A. Scott, Phys. Rev. B {\bf 66}, 064501 (2002). 
\bibitem{note1} While the scattering at ${\bf Q}=(0.53,0.53)$ and $(0.56,0.56)$ (depending on the system) may include contributions from
 multiphonon and incoherent magnetic scattering, it does not have correlated magnetic
 scattering contribution from ${\bf Q}=(0.5,0.5)$ and thus can be regarded as background scattering.

\end{thebibliography}

\end{document}